# IF:CARGO: LLM-Based Semantic Compilation for AI-Native Rule Programming Games

**Ting-Chen Hsu[1] *, Lianye Zhang[2], Jiangxu Lin[1],
Zhaoyi Yu[1], Fei Qin[3], Zihao Chen[4] ***

[1]School of Animation and Digital Arts, Communication University of China
[2]Department of Computing, The Hong Kong Polytechnic University
[3]School of Information Engineering, Lanzhou City University
[4]School of Life Science and Technology, University of Electronic Science and Technology of China

tingchenhsu.ac@gmail.com, lianye.zhang@connect.polyu.hk, chinalinjiangxu@gmail.com, jiayue5152024@gmail.com, 15209446631@163.com, the4pp1e7@163.com

## Abstract

This case study presents IF: CARGO, an experimental puzzle game that uses a large language model as a semantic compiler rather than an autonomous game-playing agent. Players author IF/THEN rules in natural language, which the model translates into a constrained command schema for deterministic validation and execution by the game engine. This architecture creates a playable loop of expression, execution, observation, and revision, framing AI interaction as semantic debugging. A mixed-methods playtest with 24 participants across eight levels examined player attempts, thinking time, perceived controllability, adjustability, and interpretations of the AI's role. Results suggest that players generally understood the model as a translation intermediary and could revise their strategies through feedback, while periodic commands, multi-robot coordination, and rule-priority mechanics created greater cognitive and diagnostic demands. The study proposes a practical pattern for AI-native gameplay: constrain natural-language input, preserve player authorship, and ensure deterministic execution.

## Introduction

Large language models (LLMs) are increasingly used in games for content generation, dialogue, puzzle generation, and agent-like interaction (Buongiorno et al. 2024; Farrokhi Maleki and Zhao 2024; Hsu et al. 2026; Kumaran, Rowe, and Lester 2024; Merino et al. 2024; Poglitsch, Szakács and Pirker 2025; Sun et al. 2025; Whitehead et al. 2025). IF: CARGO explores a different role: using an LLM to mediate player-authored rules, making natural-language interpretation itself part of gameplay. IF/THEN rule authoring has long provided an accessible form of end-user programming (Ur et al. 2014), while LLMs further enable users to express computational intent in natural language (Aveni et al. 2025; Pickering et al. 2025; Zamfirescu-Pereira et al. 2025). However, this flexibility also introduces ambiguity: players must understand how their language is interpreted (Vanderlyn, Väth and Vu 2025) and distinguish strategic errors from interpretation failures. This motivates our central question: how can natural language function as a precise, fair, and debuggable game mechanic?

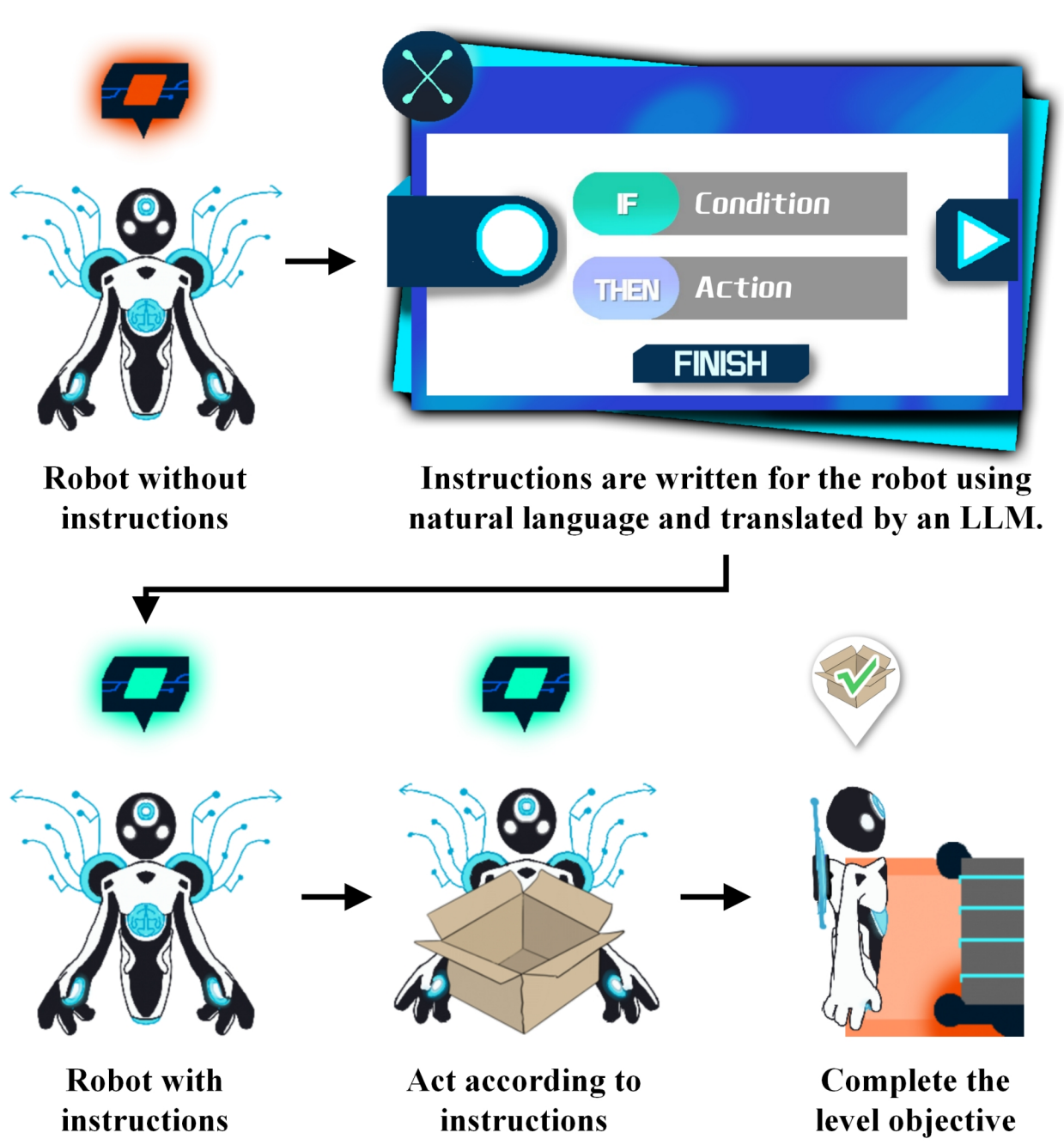

Figure 1: IF: CARGO allows players to program robot behaviors using natural language, enabling them to execute automatically within levels to complete level objectives.

We investigate this question through IF: CARGO, an AI-native puzzle game in which players program robots

* Corresponding author

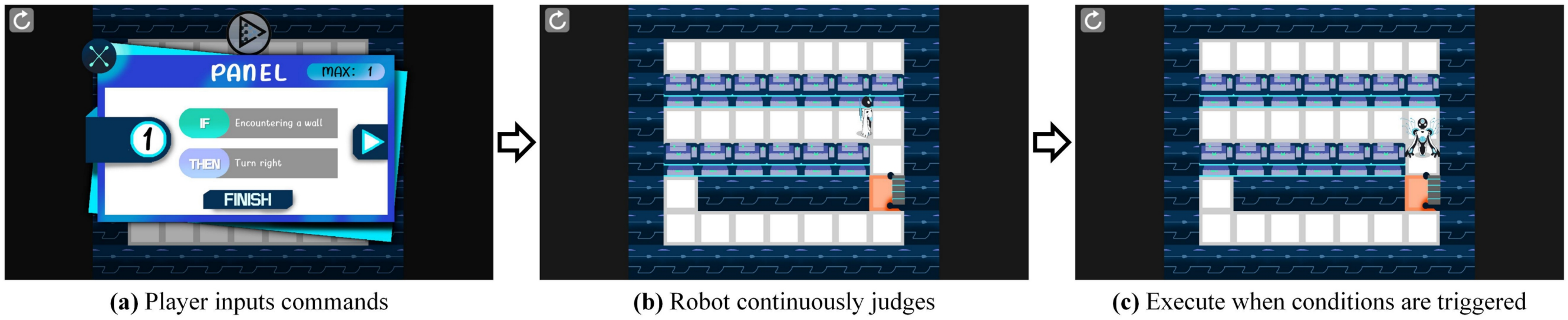

**(a)** Player inputs commands **(b)** Robot continuously judges **(c)** Execute when conditions are triggered using natural-language IF/THEN rules. The LLM acts as a semantic compiler, translating player instructions into a constrained command schema that is validated and executed deterministically by the game engine. This creates a loop of expression, execution, observation, and revision that we frame as semantic debugging. Through a mixed-methods playtest with 24 participants, we examine how players understand and revise AI-mediated rules, and derive design lessons for constrained, predictable, and playable AI-native mechanics.

Figure 2: The core mechanism and interface of the game.

## Methods

To investigate natural-language rule authoring as an AI-native game mechanic, we developed IF: CARGO around a constrained LLM-mediated execution pipeline and evaluated it through a mixed-methods playtest.

### AI-Mediated Rule Execution

IF: CARGO is a grid-based puzzle game in which players program robots through natural-language IF/THEN rules. The LLM is used only as a semantic compiler: it translates each player-authored command into a constrained internal rule representation containing a condition, an action, a target robot or chip, and optional parameters such as direction, step count, object state, or waiting behavior. The game engine then validates whether the generated rule corresponds to available mechanics and executes it through fixed simulation rules.

This architecture separates probabilistic language interpretation from deterministic game execution. The LLM does not plan solutions or control robots continuously; strategy remains authored by the player. During play, players iteratively write a rule, observe its execution, and revise either its wording or underlying strategy based on the resulting behavior.

### Playtest and Measures

We evaluated the system through a mixed-methods playtest with 24 participants across eight levels. The levels introduced reactive commands, object pickup, periodic commands, multi-robot coordination, and multi-chip rule priority.

For each level, we recorded thinking time, number of attempts, and player command input. Participants also provided two 5-point ratings measuring the perceived predictability/controllability of robot behavior and the perceived adjustability of their solution after failure. Quantitative measures were summarized using means and standard deviations. After completing the levels, participants answered a post-study questionnaire concerning their interpretation of the LLM's role, unexpected robot behavior, and limitations of the interaction. Open-ended responses were grouped into recurring themes and summarized by frequency.

## Results

Ultimately, we used Unity for prototype development and implemented the AI-driven mechanism based on ChatGPT-5.5 provided by the PlayKit platform. Figure 2 shows the core mechanism and interface of the game.

### Prototype Implementation Results

The prototype implements IF: CARGO as a grid-based puzzle system in which players program robots through natural-language IF/THEN commands. As shown in Figure 2(a), players can enter an IF instruction and a THEN instruction in the instruction panel. The left side of the panel displays the priority of the instructions. When subsequent levels require players to enter multiple sets of instructions for each robot, if multiple instructions conflict during execution, the instruction with higher priority will be executed. After writing, click the Finish button to complete the instruction embedding.

As shown in Figure 2(b), when the game starts, the robot will only perform the default forward movement behavior. However, when the IF condition in the programmed instruction is met, the robot will immediately perform the corresponding THEN behavior (as shown in Figure 2(c)). Furthermore, players can use this mechanism to write instructions that allow the IF condition to be triggered

| Level | Mechanic | Thinking time (M ± SD) | Attempts (M ± SD) | Controllability (M ± SD) | Adjustability (M ± SD) |
|---|---|---|---|---|---|
| L1 | Reactive command | 104.23 ± 72.10 | 1.42 ± 0.78 | 4.50±0.66 | 4.08±1.10 |
| L2 | Reactive command | 191.10 ± 279.63 | 3.33 ± 4.69 | 3.66±1.17 | 3.79±1.32 |
| L3 | Object picking up | 85.99 ± 57.17 | 1.67 ± 1.05 | 4.37±0.77 | 4.29±0.91 |
| L4 | Periodic command | 334.06 ± 459.81 | 4.08 ± 4.59 | 3.50±1.25 | 3.67±1.13 |
| L5 | Multi-robot | 98.65 ± 56.25 | 1.04 ± 0.20 | 4.67±0.56 | 4.54±0.66 |
| L6 | Multi-robot | 480.42 ± 411.28 | 4.04 ± 2.93 | 3.42±1.38 | 3.71±1.46 |
| L7 | Multi-chip | 421.21 ± 634.98 | 3.25 ± 3.34 | 3.54±1.53 | 3.62±1.50 |
| L8 | Multi-chip | 330.41 ± 261.63 | 3.58 ± 3.05 | 3.58±1.41 | 3.33±1.61 |

Table 1: The results of internal measurement data in the game.

| Items | Topic distribution |
|---|---|
| The LLM role perceived by players | Translation intermediary (66.7%), collaborative assistance (20.8%), instruction execution tool (12.5%) |
| Player's views on unexpected robot behavior | Self-checking logic of instructions (58.3%), rethinking the expected path (33.3%), repeated trial and error (8.3%) |
| The worst part of the experience | High difficulty and cognitive load (37.5%), sluggish response (20.8%), insufficient feedback prompt (16.7%), significant instruction limitations (16.7%), insufficient incentive (8.3%) |
| The influence of removing free input | Weakening freedom (33.3%), reducing sense of achievement (20.8%), reducing thinking burden (20.8%), reducing playability (16.7%), no significant impact (8.3%) |

Table 2: Thematic analysis of post-study questionnaire responses.

repeatedly (e.g., writing an IF instruction like "move every 10 seconds").

## Evaluation Results

The evaluation used a compact mixed-methods playtest with 24 participants. Informed consent has been obtained from all participants in this test. The study included eight levels covering reactive commands, object picking-up, periodic commands, multi-robot coordination, and multi-chip rule priority. To keep the case study focused, telemetry was limited to four core measures: thinking time per level, number of attempts, player command input, and two post-level 5-point ratings: perceived predictability/controllability of robot behavior and perceived adjustability after failure.

The results (Table 1) suggest that the interaction loop was most effective when the mapping between language, rule, and robot behavior was easy to observe. Levels L1, L3, and L5 received relatively high controllability and adjustability ratings, while L4, L6, L7, and L8 showed increased thinking time, higher attempts, and lower subjective ratings. This indicates that players could generally use natural-language rules to solve puzzles, but more complex mechanics such as multi-robot interaction and multi-chip priority placed greater diagnostic demands on them.

The post-study questionnaire (Table 2) further clarifies how players interpreted the AI's role. Most participants described the LLM as a translation intermediary rather than an autonomous player, while others saw it as collaborative assistance or an instruction execution tool. When robot behavior did not match expectations, most players first checked their own instruction logic or reconsidered the intended path, rather than immediately blaming the AI. This supports the project's central framing: IF: CARGO encourages players to treat AI-mediated behavior as something to be debugged through play.

## Discussion

Our findings suggest that the value of the LLM in IF: CARGO lies less in autonomous intelligence than in making player-authored intent executable. Most participants understood the LLM as a translation intermediary and, when behavior differed from their expectations, first reconsidered their own instructions or strategies. This indicates that constraining the model to semantic translation can preserve a relatively clear locus of player agency. This interpretation is consistent with prior work showing that users' mental models strongly shape the success and perception of interactions with intelligent systems (Vanderlyn, Väth, and Vu 2025), and with recent game research emphasizing predictability, transparency, and user agency in LLM-mediated interaction (Sun et al. 2025). Importantly, our study does not establish that this architecture causes more accurate mental models, but it suggests that a bounded AI role can make failures more amenable to player interpretation and revision.

Natural-language input, however, did not remove the cognitive demands of programming. Periodic rules, multi-robot coordination, and rule-priority mechanics required more attempts and generally received lower controllability and adjustability ratings than simpler mechanics. Natural language may reduce the need to learn formal syntax, while leaving the underlying problems of decomposition, state, timing, and rule interaction intact. This distinction is relevant to end-user programming with LLMs, where users may express computational intentions flexibly but still need to specify behavior with sufficient precision (Pickering et al. 2025). In IF, difficulty therefore emerges not only from understanding the puzzle, but from diagnosing the relationship between intended strategy, linguistic expression, compiled rule, and execution.

The responses concerning free-form input further reveal a design trade-off between expressivity and control. Some participants felt that removing free input would reduce freedom or achievement, whereas others expected it to reduce cognitive burden. This suggests that AI-native rule systems should not simply maximize linguistic freedom. Instead, natural-language expression can be paired with explicit structural feedback. For IF: CARGO, compiled-rule previews, trigger indicators, and clearer visualization of rule priority could expose how the system interprets an instruction without replacing the player's authorship. Such support aligns with broader work on LLM-supported programming that treats AI as a tool for exploring and refining user-authored solutions rather than merely producing final outputs (Zamfirescu-Pereira et al. 2025).

These findings should be interpreted within the scope of the study. The evaluation involved 24 participants, eight designed levels, and a single puzzle-game prototype, and relied primarily on descriptive measures and post-study responses. Without a comparison condition, we cannot determine whether natural-language rule authoring is more effective than conventional rule editors or other programming interfaces. The observed difficulty also combines puzzle complexity, rule-system complexity, and possible interpretation difficulty. Future studies should separate these factors and examine whether players develop more stable mental models through longer-term play.

## Conclusion

IF: CARGO demonstrates a constrained approach to AI-native gameplay in which an LLM serves as a semantic compiler rather than an autonomous controller. By separating probabilistic language interpretation from deterministic execution, the system allows players to express, observe, and revise their own strategies through natural language. Our playtest suggests that this approach can support a debuggable interaction loop while preserving player authorship, although increasingly complex rule interactions require stronger explanatory feedback. More broadly, the study suggests that meaningful AI-native mechanics need not depend on greater AI autonomy; they can instead make the interpretation and refinement of human intent itself part of play.

## Acknowledgements

This work was not funded. We thank PlayKit for its technical support in prototyping. We thank the School of Animation and Digital Arts at the Communication University of China for its support in organizing the experiment. We thank the reviewers for their feedback and all test participants for the important data they provided.

## References

Aveni, T. J.; Mor, H.; Fox, A.; and Hartmann, B. 2025. Generative Trigger-Action Programming with Ply. In *Proceedings of the 38th Annual ACM Symposium on User Interface Software and Technology*, Article 33, 1–17. New York, NY: Association for Computing Machinery. doi:10.1145/3746059.3747638.

Buongiorno, S.; Klinkert, L.; Zhuang, Z.; Chawla, T.; and Clark, C. 2024. PANGeA: Procedural Artificial Narrative Using Generative AI for Turn-Based, Role-Playing Video Games. *Proceedings of the AAAI Conference on Artificial Intelligence and Interactive Digital Entertainment* 20(1): 156–166. doi:10.1609/aiide.v20i1.31876.

Farrokhi Maleki, M.; and Zhao, R. 2024. Procedural Content Generation in Games: A Survey with Insights on Emerging LLM Integration. *Proceedings of the AAAI Conference on Artificial Intelligence and Interactive Digital Entertainment* 20(1): 167–178. doi:10.1609/aiide.v20i1.31877.

Hsu, T.-C.; Chen, W.; Lin, J.; Qin, F.; and Zhang, Z. 2026. The Double-Edged Sword of Open-Ended Interaction: How LLM-Driven NPCs Affect Players' Cognitive Load and Gaming Experience. *arXiv preprint* arXiv:2604.10107.

Kumaran, V.; Rowe, J.; and Lester, J. 2024. NarrativeGenie: Generating Narrative Beats and Dynamic Storytelling with Large Language Models. *Proceedings of the AAAI Conference on Artificial Intelligence and Interactive Digital Entertainment* 20(1): 76–86. doi:10.1609/aiide.v20i1.31868.

Merino, T.; Earle, S.; Sudhakaran, R.; Sudhakaran, S.; and Togelius, J. 2024. Making New Connections: LLMs as Puzzle Generators for the New York Times' Connections Word Game. *Proceedings of the AAAI Conference on Artificial Intelligence and Interactive Digital Entertainment* 20(1): 87–96. doi:10.1609/aiide.v20i1.31869.

Pickering, M.; Williams, H.; Gan, A.; and Ur, B. 2025. How Humans Communicate Programming Tasks in Natural Language and Implications for End-User Programming with LLMs. In *Proceedings of the 2025 CHI Conference on Human Factors in Computing Systems*, Article 875, 1–34. New York, NY: Association for Computing Machinery. doi:10.1145/3706598.3713271.

Poglitsch, C.; Szakács, F.; and Pirker, J. 2025. Evaluating Large Language Models through Communication Games: An Agent-Based Framework Using Werewolf in Unity. In *Proceedings of the 20th International Conference on the Foundations of Digital Games*, Article 25, 1–10. New York, NY: Association for Computing Machinery. doi:10.1145/3723498.3723702.

Sun, X.; Wang, L.; Li, Y.; Li, J.; Poesio, M.; Frommel, J.; Hindriks, K.; and Pei, J. 2025. Talking-to-Build: How LLM-Assisted Interface Shapes Player Performance and Experience in Minecraft. In *Proceedings of the 27th International Conference on Multimodal Interaction*, 682–692. New York, NY: Association for Computing Machinery. doi:10.1145/3716553.3756015.

Ur, B.; McManus, E.; Pak Yong Ho, M.; and Littman, M. L. 2014. Practical Trigger-Action Programming in the Smart Home. In *Proceedings of the SIGCHI Conference on Human Factors in Computing Systems*, 803–812. New York, NY: Association for Computing Machinery. doi:10.1145/2556288.2557420.

Vanderlyn, L. M.; Väth, D.; and Vu, T. 2025. Understanding the Role of Mental Models in User Interaction with an Adaptive Dialog Agent. In *Findings of the Association for Computational Linguistics: NAACL 2025*, 989–1015. Albuquerque, NM: Association for Computational Linguistics. doi:10.18653/v1/2025.findings-naacl.56.

Whitehead, J.; Wessel, T.; Chen, B.; Cruz-James, R.; Harnist, L.; Klunder, W.; Lam, J.; Lin, E.; Luo, R.; Nguyen, H.; Poddar, N.; Ravinutula, S.; Montoreano, A.; Shehane, L.; Sims, Y.; Spangler, J.; Tan, M.; and Trela, Z. 2025. Conversational Interactions with Procedural Generators Using Large Language Models. In *Proceedings of the 20th International Conference on the Foundations of Digital Games*, Article 65, 1–7. New York, NY: Association for Computing Machinery. doi:10.1145/3723498.3723788.

Zamfirescu-Pereira, J. D.; Jun, E.; Terry, M.; Yang, Q.; and Hartmann, B. 2025. Beyond Code Generation: LLM-Supported Exploration of the Program Design Space. In *Proceedings of the 2025 CHI Conference on Human Factors in Computing Systems*, Article 153, 1–17. New York, NY: Association for Computing Machinery. doi:10.1145/3706598.3714154.